\documentclass{article}

\usepackage[authoryear]{natbib}

\setcitestyle{authoryear, brackets=round}

\usepackage{amsmath}
\newcommand{\egc}{\mbox{e.\,g.\,}}
\newcommand{\vctr}[1]{\ensuremath{\mathbf{ #1 }}}
\newcommand{\ket}[1]{\ensuremath{\left|  #1 \right\rangle}}
\newcommand{\be}{\begin{equation}}
\newcommand{\ee}{\end{equation}}

\begin{document}

\title{The Case for the Everett Multiverse}
\author{David Wallace\thanks{Department of History and Philosophy of Science / Department of Philosophy, University of Pittsburgh; \texttt{david.wallace@pitt.edu}}}
\maketitle
\begin{abstract}
I give a non-technical presentation of the argument that quantum mechanics, in the form in which it is currently used, needs to be understood in many-worlds (Everettian) terms; the alternatives that have been discussed are at present not able to reproduce the full empirical content of the theory. 

This is a draft of a chapter for the forthcoming \emph{Blackwell Companion to Philosophy and the Multiverse (Klaas Kraay and Daniel Rubio, eds.) }
\end{abstract}

\section*{Introduction}

Quantum mechanics\footnote{In this article I follow contemporary physics practice by using `quantum mechanics' and `quantum theory' as synonyms. Historically, `quantum theory' was the wider term, with `quantum mechanics' referring to certain specific quantum theories; this usage is still widespread in philosophy.} was originally conceived as our best theory of microscopic matter: atoms, electrons, photons and the exotica of particle physics. But since big things are made out of small things, quantum mechanics has developed to be our best theory of pretty much the entire physical world. Quantum mechanics also pretty much entails that we live in a multiverse, and that quantum-mechanical processes are constantly causing the world to branch into unfathomably many copies of itself. 

This claim (to put it mildly) is controversial. The question of how to understand quantum mechanics is generally presented as a choice between various `interpretations of quantum mechanics', with the `Many-Worlds Interpretation' (or `Everett Interpretation') as one of a number of options that can be sequentially discussed and then compared one with another. The purpose of this chapter is to explain and defend the claim that in fact the only way to understand our best current physics is through the Everett interpretation: that is, contemporary quantum mechanics pretty much just is Everett-interpreted quantum mechanics, and there are no empirically adequate alternatives to contemporary quantum mechanics.

I begin (section \ref{section1}) with an account of the basics of quantum mechanics as applied to simple, microscopic systems --- even here, it is apparent that quantum mechanics is a radically different kind of theory from those which preceded it, and I explore this by considering the puzzle of interpreting the so-called ``quantum state''. I pursue this puzzle through a discussion of the quantum theory of measurement (section~\ref{section2}) and of modern, `unitary' quantum mechanics as it is applied in contemporary physics, including to mesoscopic and macroscopic systems (section~\ref{section3}). With modern quantum mechanics on the table, I present the quantum measurement problem as the problem of finding a consistent interpretation of that theory (section~\ref{section4}) and explain why the prospects for sidestepping that problem by replacing modern quantum mechanics are bleak (section~\ref{section5}). In section~\ref{section6} I argue that seeking a consistent interpretation pretty much leads us to a multiverse, and in section~\ref{section7} I give a very brief review of the philosophical consquences. I conclude (section~\ref{section8}) with some reflections on whether we should be using our philosophical presuppositions to decide whether to reject quantum mechanics, or vice versa.

A brief note on technical presumptions: I have tried to write this chapter so as to be intelligible to a reader with no prior physics background, but ultimately quantum mechanics is a deeply mathematical subject and arguments in this field cannot be fully developed without the mathematical details, so that reader will need to take some claims on trust. More detailed accounts of quantum mechanics aimed at philosophers can be found in, \egc, \citep{lewisontology,barrettqmbook}. If you really want to understand the subject, though, no summary for philosophers will really suffice, and you will need to study the theory as it is presented in the physics literature. There are hundreds of textbooks on quantum mechanics, at almost any desired level of difficulty and rigor, and everyone has their favorites; for myself, \citep{Dirac1930} is a classic for a reason, and I also like \citep{ballentine,cohentannoudji,peres,weinbergQM}.

\section{Quantum Mechanics for Microscopic Systems}\label{section1}

Let's start small, with a single carbon atom. Atoms have substructure, of course, but let's suppose we're interacting with this single atom at low enough energies that we can ignore its substructure and pretend it's an elementary particle. 

Classically, it's reasonably simple how to describe such a particle. The particle has a position (where it is) and a momentum (how fast it's moving --- momentum is mass times velocity), and if we know those, we know everything, because classically an `elementary particle' would be a structureless point. (And because the laws of physics are second order in time, so that both position and momentum are needed to determine the particle's future evolution.) Doing physics with this particle might consist in, e.g., preparing a beam of carbon atoms, shining that beam on a sheet with holes in it, and measuring where the atoms end up by detecting where they hit a second sheet located behind the first.

At first sight, quantum mechanics looks similar. Certainly, these sorts of experiments are standard in quantum physics: particles can be collimated into beams, and their positions and momenta can be measured, and indeed for a carbon atom, the only things that can be measured are its position, its momentum, or functions of the two. Appearances can be deceptive, though,  and a closer examination reveals profound differences.

This can be seen in a great many ways, but for our purposes the most convenient is to consider what mathematical object is used to describe the carbon atom in quantum mechanics. The answer is much more abstract than the (position,momentum) pair that would describe a classical particle: the closest analog to that description is the \emph{quantum state}.\footnote{Note to experts: in the following I am describing \emph{pure} states. This is just for expository convenience.} This rather abstract mathematical object can be described in many equivalent ways. One convenient way  is as a \emph{position wavefunction}, a complex function $\psi$ that assigns to each spatial position $\vctr{x}$ a complex number $\psi(\vctr{x})$. Another is as a \emph{momentum wavefunction}, a complex function $\tilde \psi$, not of position but of momentum, assigning a complex number $\tilde\psi(\vctr{p})$ to any momentum $\vctr{p}$. We can translate freely between these ways of writing the state (and indeed, between many other ways): as it happens, for instance, the momentum-space wavefunction is the Fourier transform of the position-space wavefunction. 

We can also refer to the state abstractly, shorn of any particular choice of how to present it: it is standard (following Paul Dirac) to write an abstract state as $\ket{\psi}$ (and to vary  the symbol $\psi$ to represent different states). States understood abstractly in this way are elements of a \emph{complex vector space}, which is to say that if $\ket{\psi_1}$ and $\ket{\psi_2}$ are elements of that space, so is $\alpha_1\ket{\psi_1}+\alpha_2 \ket{\psi_2}$, where $\alpha_1$ and $\alpha_2$ are arbitrary complex numbers. (So the complex vector space is equipped with operations of addition of vectors and of multiplication of vectors by complex numbers, which is pretty much what `complex vector space'  means. If we present the state as a function of positions or of momenta, those operations are defined in the obvious way.)

But what does it actually mean to say that a quantum system has a given quantum state? That question will occupy much of the rest of the chapter, but to get started: the quantum state plays two roles in the formalism of quantum mechanics:
\begin{description}
\item[Unitary dynamics] When a quantum system (again, like our carbon atom) is allowed to evolve over time without outside intervention, for any times $t_1,t_2$ there is a map (i.e., a function) $U(t_1,t_2)$ that transforms states at $t_1$ to states at $t_2$. That is: for a system evolving without outside intervention, its dynamics is deterministic, with the state at one time completely fixing the state at other times. This map has a mathematical property called \emph{unitarity} and so it is standard to call this dynamics the \emph{unitary dynamics}. It is normally specified indirectly, via a differential equation called the \emph{Schr\"{o}dinger equation}, but the details will not matter for my  purposes.)
\item[Born-rule probabilities] When a dynamical quantity for a quantum system is measured by an external observer, the result is given probabilistically by a rule called the \emph{Born rule}. This rule assigns to every state and every dynamical quantity a probability distribution over possible outcomes of measuring that quantity.
Special cases of the Born rule are easy to give: if the quantum state is given by the position-space wavefunction $\psi$, then the Born rule probability measure over measurements of position is 
\[\Pr(\vctr{x}|\psi)=|\psi(\vctr{x})|^2.\]
Similarly, if the quantum state is given by the momentum-space wavefunction $\tilde{\psi}$, the Born rule probability measure over measurements of momentum is
\[\Pr(\vctr{p}|\psi)=|\tilde{\psi}(\vctr{p})|^2.\]
But the rule is completely general and also assigns probability distributions to measurements of quantities like (position-squared+momentum-squared).
\end{description}

Empirically, this framework is extraordinarily successful; still, it is an odd way to present a scientific theory. For one thing, normally we think about `measurements' as processes that determine the actually-possessed values of some quantities characterizing a system, and we expect a theory to be formulated in terms of these quantities. Classical mechanics has no axioms concerning measurements, only axioms concerning the positions and momenta of bodies. For another, this description of a `quantum' theory seems very parasitic on a previous classical description: the dynamical quantities that we use to describe the carbon atom were first defined classically, and then imported into quantum mechanics. But wasn't quantum mechanics supposed to be the more fundamental theory?

We will return to both these issues shortly, but for now let's consider a third (related) oddity: a central part of my presentation was the notion of `state', but how is that notion to be understood? The most obvious possibility --- naturally suggested by the Born rule --- is that the state is \emph{epistemic}: any actual carbon atom has a true position and a true momentum, but these are unknown, and the state quantifies our partial knowledge of them. For any state $\ket{\psi}$, we might then seek a joint probability distribution $\rho_\psi(\vctr{x},\vctr{p})$ over positions and momenta, so that the Born-rule probability distribution assigned to any quantity was just the probability distribution over that quantity with respect to $\rho_\psi$. For instance,
\begin{align}
\int \mathrm{d}^3\vctr{p} \,\rho_\psi(\vctr{x},\vctr{p}) &= |\psi(\vctr{x})|^2\nonumber \\
\int \mathrm{d}^3\vctr{x} \,\rho_\psi(\vctr{x},\vctr{p}) &= |\tilde{\psi}(\vctr{p})|^2
\end{align}
and similarly for the (slightly more complicated) expressions relating the $\rho_\phi$-calculated distributions over quantities like (position-squared+momentum-squared) to their Born-rule distributions.

Quantum mechanics would then be just like classical statistical mechanics; the Born rule would simply encode the epistemic conception of the state, and reference to `measurement' in the Born rule could just be replaced by reference to the actual probability of the system having a given property; the Schr\"{o}dinger equation could just be understood as encoding how we update probabilities over time.

However, there is no such $\rho_\psi$, and the epistemic conception is untenable. A variety of no-go theorems\footnote{The earliest is due to \citet{vonneumann}; more powerful results were proved by \citet{gleason} and by Bell, Kochen, and Specker~\citep{bell1966,kochenspecker}. See \citep{redheadbook} for discussion.} establish rigorously that this kind of epistemic approach is untenable, but the underlying reason is easy to understand: quantum theory displays a phenomenon called \emph{interference} that is incompatible with the epistemic interpretation. To understand this, consider the famous `two-slit experiment', in which a beam of (say) carbon atoms is fired at a barrier pierced by two thin slits, and a screen is placed on the far side of the barrier, and the pattern of impacts of atoms on the screen is reported. That pattern is a measurement of the probability a given atom has of impacting at each point on the screen, and if the epistemic interpretation were correct, the probability of the atom reaching point $X$ would be the sum of its probability of reaching $X$ via slit 1 and via slit 2. And it is not. Any introductory course on quantum mechanics will explain the details, but in essence: 
\begin{itemize}
\item The Born rule gives the probabilities of the atom reaching each screen in terms of the mod-squared amplitude of the position wavefunction $\psi$: 
\be\Pr(\vctr{x})=|\psi(\vctr{x})|^2.\ee
\item That wavefunction can be decomposed as $\psi= \psi_1+\psi_2$, where $\psi_i$ (up to normalization) is the wavefunction that would have resulted if the particle had been forced to pass through slot $i$ (say, by covering up the other slot).
\item The probability distributions corresponding to the particle reaching the screen through slot $i$ is the mod-squared amplitude of $\psi_i$: 
\be\mathrm{Pr}_i(\vctr{x})=|\psi_i(\vctr{x})|^2.\ee
\item By simple calculation, 
\begin{align}
\Pr(\vctr{x})&= |\psi_1(\vctr{x})+\psi_2(\vctr{x})|^2\nonumber \\
&=|\psi_1(\vctr{x})|^2+|\psi_2(\vctr{x})|^2 + (\psi_1^*(\vctr{x})\psi_2(\vctr{x})+\psi_1(\vctr{x})\psi_2^*(\vctr{x})).
\end{align}
This equals $\Pr_1(\vctr{x})+\Pr_2(\vctr{x})$ only if the `cross term' $(\psi_1^*(\vctr{x})\psi_2(\vctr{x})+\psi_1(\vctr{x})\psi_2^*(\vctr{x}))$ vanishes, which in general it does not: the total probability need not be the sum of the separate probabilities but might be greater (constructive interference) or less (destructive interference).
\end{itemize}

So much for the epistemic conception.\footnote{Note to experts: the broader class of `$\psi$-epistemic' approaches (as explored in, e.g., the ontic models framework; cf \citep{spekkensepistemic}) would largely be classified as `inferential' on the taxonomy I use here; by `epistemic' I mean the narrower class of approaches where the ontic state has definite values of all the quantum observables. } There are two other natural possibilities to consider. Firstly, the state might be \emph{representational}: that is, it might be an encoding of the various physical properties the system (in this case, the carbon atom) actually has, and distinct states for a given system correspond to distinct ways that system could be. `States' in nonquantum physics are normally representational: the classical state of a carbon atom, for instance, is a point in a six-dimensional space coordinatized by three components of position and three components of momentum (equivalently, it's the ordered pair $(\vctr{x},\vctr{p})$ of the atom's position and momentum). So in some ways this is the most natural choice. 

But it is mysterious both what physical properties are being represented, and how to make sense of the Born rule, on this approach. A typical wavefunction, for instance, doesn't represent a carbon atom as having any particular position; the nearest we can say seems to be that it is indefinite between many positions. And yet according to the Born rule --- and according to experiment --- position measurements on a system in that indefinite state invariably return definite results.

The other possibility is that the state is \emph{inferential}: that is, it should be understood directly as providing advice to the experimenter as to what outcomes to expect on carrying out a measurement, without any mediating account in terms of the physical properties being measured. This notion of state is not really found elsewhere in physics\footnote{Arguably the Jaynesian conception of statistical mechanics comes close; cf \citep{jaynesstatmech,jaynesstatmech2}.} but in philosophy it bears a strong resemblence to the logical-positivist idea that `theory terms' should be interpreted not as literal descriptions of the unobservable world but as a calculus to predict the results of experiments.

The inferential conception of the state, like the epistemic conception, makes natural sense of the Born rule: the advice given by the state about measurements of a given quantity is precisely the probability distribution it defines over 
that quantity (and the unitary dynamics just tells us how to update that advice over time). And, precisely because it is silent about the real physical properties of quantum systems, it is not committed to classically indefinite values of those quantities. 

Yet for that very reason the inferential conception is radically incomplete. What actually makes it true that a given piece of apparatus in the lab is a device for measuring the positions of carbon atoms? The naive answer would be that the device is built so that some degree of freedom of the device becomes correlated with the actual position of the atom. But on the inferential conception, quantum mechanics is not in the business of describing the `actual position' of the atom, only what we should expect if we measure it, and so a vicious circularity beckons. To see how this plays out, let's consider what actually happens when we try to extend quantum mechanics to discuss the measurement process itself.

\section{The quantum theory of measurement}\label{section2}

Most scientific theories don't attempt to model the very process by which measurements are made of those theory's quantities. Zoology does not tell us how to count animal populations; celestial mechanics contains no theory of optics. Often, this is because the quantities those theories describe are directly observable, but even when this is not the case, there is no general requirement that the theory's empirical domain includes its own measurement devices. However, if those devices are not unaided human senses, and are not simply found by chance but actively constructed by human expertise, some theory must describe their workings. The workings of the devices that detect gravitational waves are not describable within the mathematical formalism of general relativity, but they must be, and are, describable within the formalism of \emph{some} theory --- they are not naturally occurring, and were not discovered by chance.

So too for quantum mechanics. The original subject matter of quantum theory --- atoms and subatomic particles --- is paradigmatically unobservable by the unaided senses. Either quantum mechanics itself describes the workings of devices that measure quantum properties, or else some other theory does. 

At the inception of quantum mechanics, it was possible to accept this second possibility, and indeed to take that `other theory' to be \emph{classical} mechanics. The principles appealed to in the working of the first quantum experiments were, by and large and for the most part, classical principles (how could it be otherwise, since quantum principles were only in the process of being discovered). And even as the laws of nature governing classical quantities like position and momentum became modified, with the Schr\"{o}dinger equation increasingly supplanting classical equations of motion, still it was possible for a time to suppose that those classical quantities exhausted the content of quantum mechanics. (Note that in my discussion of the carbon atom I assumed exactly that, appealing to the classical notions of position and momentum to explain the quantum axioms.)

It is instructive to imagine a possible world in which this conception of quantum mechanics persisted indefinitely. In a world like that, every physical system would have a classical description, and some, perhaps all, would also have a quantum description. That quantum description would contain nothing ontologically novel: the quantities it posited would simply be classical quantities. In this picture of quantum mechanics, the classical world might be \emph{dynamically} dependent on the quantum world, in that at the microscopic level the classical equations would be at most approximately correct and accurate predictions would require quantum mechanics. But, symmetrically, the quantum world would be \emph{conceptually} dependent on the classical world, with classical mechanics supplying the physical quantities and underpinning our understanding of the ontology.

This picture of quantum mechanics would fit naturally with the inferential conception of the quantum state, and would resolve the problems for that conception that I identified previously. What are quantum inferences about? --- classical quantities. And how are we to understand the measurement of those quantities? --- ultimately, classically. I do not intend this as a historical essay, but it is worth noting that this way of thinking about quantum mechanics tracks at least some strands of the so-called `Copenhagen interpretation' of quantum mechanics. (See, \egc, \citep{saunderscopenhagen} for a presentation of the Copenhagen interpretation along roughly these lines.)

Is this picture philosophically coherent? It's a delicate point. Certainly it's not a scientific-realist theory in any straightforward sense: quantum theory is to be understood inferentially and not dynamically; the relation between the classical and quantum equations is subtle at best. (Earlier texts on quantum mechanics often referred to a `correspondence principle' connecting the classical and the quantum, but the interpretation of that principle is notoriously obscure.) But it is also a moot point, because the possible world in which quantum mechanics retained this permanent dependence on classical mechanics is not ours. In our world, it has long since proved inadequate, for two (related) reasons.

The first reason is that our description of the microscopic world long outgrew the confines of classical physics; indeed, it began to do so almost at the dawn of quantum theory. I chose carbon atoms advisedly in my previous example: most atoms --- and all stable or long-lived subatomic particles --- have an additional property called \emph{spin}. Spin is \emph{something like} angular momentum; to say that a particle has a certain spin is \emph{something like} saying that it is spinning on its axis at a certain rate. But it's not \emph{that much} like that; the rotational interpretation of spin is ultimately a metaphor or analogy, a ladder that students climb when learning quantum theory but swiftly kick away.

Spin is already outside the descriptive resources of classical physics, but it was only the beginning: modern quantum theory is replete with more examples. To pick two: the very notion of \emph{particle}, once we get to relativistic physics, breaks some of its connections to the classical notion of particle; notably, its position is definable at most in an approximate sense. And a proton is \emph{in some sense} made out of three quarks, but that sense bears only a little relation to classical notions of mereology: the compositional story of quarks is a metaphor at best, and a looser metaphor even than spin.\footnote{See \citet{wallace-leedsrealism} for a somewhat more detailed account of a third example, the quark/gluon plasma of the early universe. See also \citet{leightonqmpragmatism} for a heroic attempt to corrall some of these quantum concepts into a classical framework.}

The second reason is that it long since became necessary to appeal to quantum-mechanical principles in the design of experiments. The workhorse of modern atomic physics is the laser; but LASER is an acronym for Light Amplified by Stimulated Emission of Radiation, and stimulated emission is an inherently quantum phenomenon. Many more examples exist, and indeed apply in experimental science more broadly: gravity wave astronomy, and the positron emission tomography increasingly used in medical imaging, are both inherently quantum.

Modern quantum theory --- not (only) for \emph{a priori} reasons but simply as an empirical observation --- is a theory which models its own measurement processes and which contains classical mechanics as an approximate, emergent limit, not as an ineliminable co-star. Let's now see what that modern theory looks like.

\section{Unitary quantum mechanics}\label{section3}

A full statement of the structure of modern quantum mechanics --- or `unitary quantum mechanics' as it is sometimes called\footnote{'Everettian minimalism' might be another suitable name.} --- is far beyond the scope of this chapter. It will suffice to list what I take to be its key features, at least as far as the interpretation of quantum mechanics is concerned.
\begin{description}
\item[Unitarity] All closed systems, even meso- or macroscopic ones, are quantum systems that evolve unitarily under the Schr\"{o}dinger equation.
\item[Dilation] The effects of some outside system (such as a measurement device) on a quantum system can always be modelled by including the outside system as another quantum system and describing the evolution of the combined system unitarily. (In the quantum information community this principle is sometimes referred to as the Church of the Larger Hilbert Space.\footnote{Note for UK readers: As we will see, the Church of the Larger Hilbert Space stands to the Many-Worlds Interpretation roughly as the Church of England stands to the Roman Catholic church.})
\item[Emergent Classicality] The validity of a classical description of a system can always be understood as a certain approximation to a quantum description of that same system (and can always, in principle, be set aside in favor of that quantum description).  In each case, the degrees of freedom that obey approximate classical equations are coarse-grained or collective degrees of freedom of the underlying quantum system, such as the center-of-mass degree of freedom of a dust mote or sand grain, or the local density of a fluid of particles.
\item[Decoherence] Quantum interference is suppressed for the dynamics of collective degrees of freedom like these, usually through the mechanism of redundantly recording them in microscopic degrees of freedom. This means that with respect to these degrees of freedom, the Born-rule probabilities evolve over time like normal probabilities; they do not display the reinforcing or cancelling out that characterize interference in microscopic quantum systems. This often means that the dynamics of mesoscopic degrees of freedom are (formally speaking) stochastic rather than deterministic, as in (e.g.) the Brownian motion of a pollen grain; the `classical regime' of quantum mechanics is taken to include stochastic processes like this as well as deterministic processes like Newtonian mechanics.

Conversely, the presence of decoherence with respect to certain collective degrees of freedom (along with the existence of robustly autonomous dynamics for those degrees of freedom) can be taken as an indicator of when classical mechanics is emergent. (There is no automatic rule that any old collective degree of freedom has these properties.)

\item[No special role for measurement] ``Measurements'' and ``observations'' are just dynamical processes coupling one system with another; these terms have no special role either in the formalism or the interpretation of the theory.

In particular, the reference to `measurement' in the statement of the Born Rule in practice drops away. It is replaced by the requirement that the Born Rule is only applied to decoherent degrees of freedom. Because interference is suppressed for decoherent degrees of freedom, if the Born rule is only applied in this situation then no contradictions arise. 
\end{description}

To be clear: I am attempting here to be descriptive, not normative. I am not saying that we \emph{should} formalize quantum mechanics in accordance with these five features; I am saying that \emph{in fact} modern quantum mechanics conforms to these five features. Unitary quantum mechanics is the framework used in  contemporary physics whenever detailed considerations of measurement processes, observation, or the quantum/classical transition arise. It is obviously somewhat difficult to demonstrate this rigorously --- and there is always some element of rational reconstruction involved in extracting the underlying principles of an area of science from the tangle of practice. So for the most part I will simply ask readers to take on trust that this is an accurate description of contemporary quantum mechanics. If you are not so trusting, the fallback is to wrap the rest of the chapter in a conditional hypotheses: supposing that unitary quantum mechanics is an accurate formulation of contemporary quantum physics, what follows?

I should, however, address one (supposed) feature of quantum mechanics that is not part of unitary quantum mechanics as I have described it: the so-called \emph{collapse of the wave-function}, also called the \emph{projection postulate}. This is the idea that when a  measurement is performed, the wavefunction is replaced by a new one, determined probabilistically via the Born rule. If, for instance, the position-space wavefunction of an atom is 
\be
\psi(\vctr{x})=\alpha \psi_L(\vctr{x})+\beta\psi_R(\vctr{x})
\ee
where $\psi_L:$ and $\psi_R$, respectively, are localized in the left and right halves of some box, and then a measurement is performed to determine which side of the box the atom is on, the Born rule says that the probability of the measurement giving the result `Left' is $|\alpha|^2$ and the probability of it giving `Right' is $|\beta|^2$. The projection postulate says in addition that if the result is `Left', the quantum state should be replaced with $\psi_L$; if it is `Right', it should be replaced with $\psi_R$.

In the context of unitary quantum mechanics, this is ambiguous between two ideas: one innocuous, one radical. To see this, recall that in unitary quantum mechanics measurement is a dynamical process, which schematically will look something like this:
\be\label{quantum-measurement}
(\alpha\ket{\psi_L}+\beta\ket{\psi_R})\otimes\ket{\text{``Ready''}}\rightarrow \alpha\ket{\text{``Outcome is L''}}+\beta\ket{\text{``Outcome is R''}}
\ee
where $\ket{\text{``Ready''}}$ is some state of the measurement device and its environment in which the device is ready to function, and $\ket{\text{``Outcome is L''}}$ and $\ket{\text{``Outcome is R''}}$ are joint states of the atom, the measurement device, and its environment in which (respectively) the device reads `L' and `R'. Given decoherence, there will be no subsequent interference between these two terms in the state, and so if we want to use the Born rule to calculate, say, the probability of a subsequent measurement of position giving result `L' conditional on the first result giving `L', it will be harmless to do so by using the new quantum state $\ket{\text{``Outcome is L''}}$ in place of the original superposition. (Harmless, but optional: if we just choose to keep the full state and then extract conditional probabilities from joint probabilities, the result will be the same.) 

So: as long as we apply it only to degrees of freedom which are decohered, the collapse rule can be considered an innocuous calculational shortcut. Where it becomes radical is if we suppose that state collapse is some \emph{dynamical} process that sometimes replaces unitary evolution, and that can apply even to not-fully-decohered degrees of freedom.  A collapse law of this kind would make different empirical predictions to unitary quantum mechanics, predictions that would in principle be measureable through sufficiently careful interference experiments. To date there is no evidence whatever for any such dynamical collapse (I return to this issue in section~\ref{section5}). In addition (as I discuss in more detail in \citep{wallaceorthodoxy}), taking the collapse-on-measurement rule too literally leads to paradox or contradiction in many standard experimental contexts, such as relativistic quantum mechanics or continuous measurement.

Both \citet{Dirac1930} and \citet{vonneumann} include versions of the collapse rule in their classic textbooks, and many textbooks on quantum mechanics still include some version of it. Mostly these books are ambiguous as to whether it should be understood in the innocuous or radical sense. (Dirac seems to have had the radical sense in mind; von Neumann is more guarded.) Textbooks that use rather than present quantum mechanics pretty uniformly omit any mention of collapse (I know of no quantum field theory textbook that discusses collapse, for instance) and it is not invariably found even in modern quantum textbooks: \citep{ballentine} and \citep{weinbergQM} are two examples by eminent scholars. (Weinberg does mention (but not endorse) collapse, but only in a parenthetical section on interpretation; it plays no role in his development of the theory.)

I discuss the status of collapse in modern physics in more detail in \citep{wallaceorthodoxy}; from here on I will continue to assume that it exists in at most its innocuous and optional form and that the dynamics of modern quantum mechanics is unitary.

Before proceeding, I should note that I refer to `modern' quantum mechanics advisedly. Older work in quantum mechanics made much wider use of quantum-classical correspondences, of explicit discussion of measurement and observer, and the like, and often handled the quantum/classical transition in confused and confusing fashion. The version I present here coalesced in the 1980s and 1990s with the development of decoherence theory and quantum information, though to the best of my knowledge there is no systematic historical study of this transition.

\section{Interpreting unitary quantum mechanics}\label{section4}

Now that we have a clearer idea of what contemporary quantum mechanics actually looks like, let's return to the question of how to interpret the quantum state, and reconsider the three possible conceptions of the state I introduced in section \ref{section1}: inferential, epistemic, representational. 

The difficulty for the inferential conception should already be apparent from our previous discussion: what are these inferences \emph{about}? They cannot be about measurement outcomes understood primitively, because `measurement' is not a primitive in modern quantum mechanics (and because we need it not to be a primitive in order to make sense of our successful science of quantum measurement devices). They cannot be about classical facts, because `classical' is also not a primitive in modern quantum mechanics: classical physics is just an emergent, approximate regime of quantum physics, and we cannot reintroduce it as primitive on pain of empirical inadequacy. 

Philosophers of science have seen this movie before.\footnote{I develop this point in more detail in \citep{wallace-leedsrealism}.} As I mentioned in section~\ref{section1}, the inferential reading of quantum mechanics is a close relative of the logical-positivist and logical-empiricist programs of the early twentieth century, in which `theory terms' in scientific theories were not to be taken literally as claims about unobservable reality, but were to be understood as inference tickets legitimating claims made in a supposed observation language. And what ultimately led to those theories being judged untenable even in their own terms\footnote{See, e.g., \citep[ch.1]{vanfraassenscientificimage} for a version of this criticism of logical empiricism} was the realization that observation is theory laden: that there is no way to draw a distinction between observation and theory clean enough to allow theory claims to be inferential tools for deriving observation claims. 

So too for the inferential reading of the quantum state. Until and unless we are given a clean account of what the inferences are inferences about --- and no such account seems to be available in modern quantum mechanics, which is quantum through and through  --- it does not provide a viable way to make sense of quantum mechanics.

(We can see this at work in the most widely discussed inferentialist approach to interpreting quantum mechanics: QBism. (See, \egc, \citep{fuchsmerminschack,fuchsschackgreeks}.) QBists interpret quantum states as encodings of agents' beliefs about the result of observations ---but this requires them to treat `observation' as a primitive, and largely disconnects their approach from any concrete application of quantum theory. As such, the approach has yielded interesting results in the abstract domain of quantum information, but at present remains largely silent on how to understand quantum mechanics \emph{as applied to concrete physical systems}. QBism, for instance, at present lacks the resources to make sense of standard quantum-mechanical predictions like the excitation spectrum of hydrogen or the prediction that helium-3 is superfluid at a lower temperature than helium-4, or of canonical experiments like the measurement of the charges and masses of particles (which on the representational reading are parameters in a dynamical equation for how a system's objective features evolve over time) and it is entirely unclear how it could be modified to add them. See \citep{wallace-leedsrealism} for more detail on this point.)

As for the epistemic conception of the state: we saw that this was untenable at the microscopic level, and this is not alleviated simply by considering still-larger systems. It is true, and highly suggestive, that no contradiction arises if we adopt an epistemic attitude to decohered variables. But decoherence is an approximate and emergent concept, and does not in itself appear to give us the resources to make sense of quantum theory as a whole. Furthermore, even a very complicated superposition could in principle be recohered by a sufficiently skilled and resourceful outside agent, and if we were to adopt an epistemic reading of that superposition we would again run into conflicts with interference phenomena.

That seems to leave the representational conception of the state as the only remaining option, but this too runs into severe problems. We saw that at the microscopic level, the representational approach seems to force some kind of indefiniteness: superpositions of particles in different positions appear to be in no definite position at all. This indefiniteness persists into the macroscopic world when we consider the quantum theory of measurement: if we measure the position of a particle whose position is indefinite, the result is that the measurement device itself gives an indefinite outcome. Consider again equation (\ref{quantum-measurement}), for instance: the final state of the combined system of atom, device and environment is a superposition of a state representing outcome `L' and a macroscopically different state representing outcome `R', whereas experience tells us that the outcome should be either `L' or `R', with some probability of each. 

We can make this vivid via Schr\"{o}dinger's famous thought-experiment: suppose that our experimental setup ``records'' the result of measuring the particle to be on the left by killing a housecat; if the measurement finds the particle on the right, the cat lives. Then if before the experiment the particle's position is indefinite, after the experiment it is indefinite whether the cat is alive or dead, in (apparent) contradiction with our experiences and with the Born rule. (Note that by contrast, the epistemic reading of the quantum state does just fine with Schr\"{o}dinger's cat: decoherence prevents any reinterference of live and dead cats, and the superposition is interpreted not as a weird indefinite cat but mundanely as ignorance as to whether the cat is alive or not.)

This is the quantum measurement problem in modern quantum theory.\footnote{In general in philosophy, disagreements between experts are as much about what the questions are as about how they should be answered. The version of the measurement problem I present here is by no means universally adopted; other versions identify `the measurement problem' with the problem of when wavefunction collapse occurs (which presupposes that collapse is part of the quantum formalism), or with the fact that quantum mechanics leads to indefinite outcomes unless modified or supplemented (which presupposes a representational conception of the quantum state). I explicate and defend the formulation I give here in \citep{wallaceorthodoxy}, } We want to understand the content of the theory; at the macroscopic level the only option seems to be to regard it as epistemic; at the microscopic level the only option seems to be to regard it as representational. There appears to be no consistent way forward.

It is important to appreciate that the measurement problem, in this sense, does not really get in the way of scientific practice. Unitary quantum mechanics is perfectly sharply defined mathematically and provides entirely successful advice as to how to make predictions in any physical circumstance we can name (as was arguably not the case in the earlier versions of quantum theory that appealed to quantum-classical correspondence and/or to observers and measurements as primitive). The best rational reconstruction I know of the tacit interpretative strategy of those physicists unconcerned about the measurement problem is just that they read the quantum state representationally when decoherence is not present and epistemically when it is, and do not trouble themselves with the contradiction. Physicists sometimes speak of the `shut up and calculate' interpretation, and this is basically what it consists of, but it is not really a consistent interpretation but a quietist choice not to seek one. (Put another way: the `shut up and calculate interpretation' makes two demands of its adherents. One of those two demands is that they calculate.)

Still, if we want to \emph{understand} quantum mechanics, that would in turn seem to require a consistent way to understand the theory; and it seems that there is none to be had. Should this shake our faith in quantum mechanics itself?

\section{Change the theory?}\label{section5}

If there is no consistent way to understand quantum mechanics, perhaps that means that quantum mechanics is \emph{wrong}, or at the least \emph{incomplete}, as an account of the physical systems it purports to describe. From that point of view, the quantum state would be inferential in the sense that the predictions we make from it are by and large empirically correct. But a correct understanding of why they are correct would come not from interpreting and understanding quantum mechanics, but by completing, modifying, or outright replacing it.

This idea came to dominate the philosophy of physics in the 1990s and remains popular today, at least among philosophers. And many, many such completions, modifications and replacements have been proposed. The most popular have been:
\begin{description}
\item[$\psi$-epistemic hidden variable theories:] these have something of the flavor of the epistemic conception of the quantum state that we have already considered and rejected, but now the quantum state is supposed to encode probabilities not over classical observables but over some other `hidden' variables, so that the true physics is a dynamical theory of these variables and quantum theory is recovered as a statistical-mechanical overlay on that dynamical theory. A variety of no-go theorems, most famously the PBR theorem (\citet{puseyetal}) and Bell's theorem \citep{bell1964,bellsocks} tell us that these variables, if they exist, must be exotic and alien in a variety of ways; see, \egc, \citep{Leifer2014} for further discussion.
\item[Pilot-wave hidden variable theories:] these retain the quantum state as part of the basic formalism, but supplement it with additional variables, again usually called `hidden', such that either the state and the hidden variables jointly, or the hidden variables alone, represent physical goings-on. The hidden variables have their own dynamics but that dynamics depends on the quantum state. The best known is the de Broglie-Bohm theory, aka Bohmian mechanics (\citealp{bohmI,bohmII}; see, \egc, \citealp{goldsteinsep} for a presentation).
\item[Dynamical-collapse theories:] At the microscopic level the quantum state cannot be treated epistemically but at the macroscopic level it seems that it must be; and this might suggest that we need some \emph{dynamical transition} between these two readings of the quantum state. The resultant theory would retain a representational reading of the quantum state, but modify the unitary dynamics, replacing the Schr\"{o}dinger equation with some partially-random dynamical process.

Collapse of the wavefunction basically achieves that, if we replace `collapse upon observation' with `collapse upon irreversible decoherence', but both observation and decoherence are emergent, higher-level notions unsuitable for playing a fundamental role in a theory of physics. Dynamical collapse theories try to replace `collapse upon measurement / decoherence' with something like `collapse upon these precisely-defined conditions', where those conditions are as far as possible chosen to be roughly coextensive with decoherence. (A particularly popular approach, advocated \emph{inter alia} by \citet{diosi1987} and \citet{penrose-1996}, would tie collapse to gravitational effects, and thus would aspire to connect to the search for a quantum theory of gravity.) Since there is no prospect of \emph{exact} coextensivity, this means that theories of this kind can in principle be tested, and indeed many versions of these theories have already been ruled out empirically and the entire class of theories like this may well be disconfirmed --- or triumphantly vindicated! --- in the next few decades.\footnote{For a general review of dynamical collapse theories and their empirical status, see \citep{bassi-dorato-ulbricht}; for philosophical discussion of the proposed tests of Diosi-Penrose style theories, see \citet{huggettgravitylaboratory}.} (By contrast, the two kinds of hidden variable theories are usually constructed so as to be empirically equivalent to unitary quantum mechanics.)
\end{description}

The natural approach to these alternatives to quantum mechanics, at least for a philosopher, would be to start tabulating their conceptual strengths and weaknesses, so as to begin to consider which replacement of quantum mechanics to adopt. And indeed, many such tabulations can be found.\footnote{My own is \citep{wallaceashgate}; other recent accounts include \citep{barrettqmbook,lewisontology,maudlinqmbook}.} But this would be to miss a far more important fact about strategies of this kind:

\textbf{No known modification, completion, or replacement of quantum mechanics exists that reproduces the full range (or even very much of the range) of unitary quantum mechanics.}

To be slightly more specific: if we are interested onlty in the interactions of nonrelativistic particles in the absence of electromagnetic radiation, then there exist some alternative theories, including dynamical-collapse theories (the GRW and CSL theories) and pilot-wave hidden variable theories (the de Broglie-Bohm theory) that can reproduce the predictions of quantum mechanics in that domain. But these are toy theories (proofs of concept, if you like.) The great bulk of quantum predictions rely to one degree or another on \emph{quantum field theory} and (notwithstanding some interesting preliminary work) there are no satisfactory extensions of any of these approaches to quantum field theory that suffice to reproduce its predictions. (I expand on this claim in \citealp{wallacebluesky}.)

Could any such theory be found, with sufficient effort? I have argued (again, in \citealp{wallacebluesky}) that there are reasons to be skeptical. But whatever the virtues or vices of those reasons, seeking a modification or completion of quantum theory that is empirically equivalent to the \emph{full} scope of contemporary unitary quantum theory is a (profoundly difficult) scientific research program, not a program of philosophical interpretation. 

Viewed from a great distance, the problem for change-the-theory approaches is the same as the problem for inferential conceptions of the quantum state (like QBism). The problem with these approaches is not internal to them; I don't have a conceptual problem with, say, adopting an inferential reading of some part of physics, or exploring modifications of the Schr\"{o}dinger equation. The problem is that no extant way of doing so actually succeeds in making the myriad confirmed predictions about concrete physical systems that unitary quantum mechanics makes.

\section{The road to Everett}\label{section6}

If we want to understand our extant physics rather than (or as well as) contributing to the search for possible changes to extant physics, unitary quantum mechanics is the only game in town. But haven't we seen that there is no stable way to understand unitary quantum mechanics, so that the best we can do is to shift inchoately between a representational reading of the theory at the micro level and an epistemic reading once decoherence has set in?

As \citet{everett} first pointed out, though, this is too quick. Recall what is apparently wrong with the representational conception of the quantum state: it predicts that macroscopic systems (like measurement devices, or cats) end up in indefinite states, and we do not see them as in indefinite states. But should we expect to see them? In fact, quantum mechanics says otherwise.\footnote{The argument that follows is adapted from \citep[ch.6]{wallacevsi}}

In unitary quantum mechanics, a human observer like me is just one more physical system. If I look at Schr\"{o}dinger's cat, then very schematically I must have at least three relevantly distinct states, which we might write as $\ket{\text{Ignorant}}$ (the state I am in before I see the cat), $\ket{\text{Sees Alive}}$ (the state I am in once I see that the cat is alive), and $\ket{\text{Sees Dead}}$ (the state I am in once I see that the cat is dead).

Suppose I look at a definitely-living cat; before the observation, the joint state of cat and me will be $\ket{\text{Alive; Ignorant}}$, and this state will evolve\footnote{Technically, the cat will evolve itself into another `alive' state, but we can ignore this for simplicity} into $\ket{\text{Alive; Sees Alive}}$:
\be
\ket{\text{Alive; Ignorant}}\rightarrow \ket{\text{Alive; Sees Alive}}.
\ee
Similarly, if the cat is definitely dead to start with, the process of observation must proceed like this:
\be
\ket{\text{Dead; Ignorant}}\rightarrow \ket{\text{Dead; Sees Dead}}.
\ee
Now suppose the cat starts in the Schr\"{o}dinger-cat state 
\be
\ket{\text{Cat State}}=\alpha\ket{\text{Alive}}+\beta\ket{\text{Dead}}.
\ee
 Intuitively, we might expect my observation of this system to look something like
\be
\ket{\text{Cat State; Ignorant}}\rightarrow \ket{\text{Cat State; Sees Weird Indefinite Cat}}.
\ee
But intuition is a poor guide to physics, and what the physics actually tells us (as an automatic consequence of how observations of definitely-living and definitely-dead cats go) is that \ket{\text{Cat State; Ignorant}} can be rewritten as
\be
\ket{\text{Cat State; Ignorant}}=\alpha \ket{\text{Alive; Ignorant}}+\beta \ket{\text{Dead; Ignorant}}
\ee
and so it evolves like this: 
\be
\ket{\text{Cat State; Ignorant}}\rightarrow \alpha\ket{\text{Alive; Sees Alive}}+\beta\ket{\text{Dead; Sees Dead}}.\ee
According to quantum mechanics, I do not evolve into a state of definitely seeing an indefinite cat; I evolve into an indefinite state of my own, a state which is at a superposition of two ordinary, macroscopically definite, states.

And this goes on. If you ask me whether the cat lives, you end up in two states at once: one of hearing me say “yes”, one of hearing me say “no”. Indeed, the combined state of all of us – you, me, and the cat – is a superposition of two states at once, but both individual states are boringly ordinary: the state where we come across a live cat, or the state where we come across a dead cat. If a third person enquires, or if you post the cat’s status on social media, more and more systems get drawn into this superposition.

Indeed, above a certain scale – a scale much smaller than the poor cat – interaction between one system and another is unavoidable even if there is no intentional “looking”. The gravitational effect of the cat on the air around me or the particles in my body effectively entangles me, and you, and our surroundings, with the cat whether or not we attempt to know its condition. Try to put anything the size of a cat into a superposition of dead and alive, and before long the whole planet will be in a superposition of classically definite macroscopic states. Each by itself is a normal, macroscopically (approximately) definite, state of the Earth; the two differ only in whether some unfortunate cat lived or died. Each one develops in time according to the ordinary rules that govern normal states of the Earth. The state of the Earth, that is, consists of two parallel branches: a “cat lived” branch, and a “cat died” branch, each evolving over time without reference to the other. (And because of decoherence, they will not interfere with one another.)

There is a good word for a part of reality that looks like the ordinary Earth and which evolves without reference to other parts of reality: a world. Not a world in the sense of an entire self-contained universe, but a world in the sense that Earth or Mars is a world: they are pieces of reality that interact strongly with themselves but are scarcely affected by one another. 

Of course, experiments with cats are scarcely the only place where the effects of quantum theory magnify up to human-sized objects. We live in a world where small changes at the micro level can, over time, reach the scale of the everyday. An electron in a fluorescent light is both here and there, a cosmic ray both does, and does not, strike a DNA chain in a cell… and, soon enough, the light both does and does not flicker; the cell both does and does not mutate. So this splitting into parallel branches is commonplace, occurring countless times in a second, all across the Earth.

We are led to the conclusion that in fact we can make sense of unitary quantum mechanics via a representational reading of the quantum state, without any conflict between the theory and our experiences. But to do so is to accept that the world we live in is one of an innumerably greater plurality – an emergent multiverse - all existing in parallel with one another, each one constantly branching from the others. Hence the better-known name for the Everett interpretation of quantum mechanics: the many-worlds interpretation.

\section{Philosophical consequences of quantum theory}\label{section7}

We have seen that if our best current theory of physics is correct, we live in a branching multiverse, whose detailed structure is given by that theory. The metaphysical and epistemic consequences are legion; here I mention some of the most important. (In each case I expand substantially on these mentions in \citep{wallacebook}.)

\subsection*{Emergent ontology}

Quantum mechanics tells us that apparently bedrock concepts like the spatial location of bodies are at best emergent and approximate. The microscopic structure of the world is represented by the alien mathematics of quantum theory and may not be describable at all, except approximately and metaphorically, in the familiar object/property language of metaphysics~\citep{wallacemathfirst}; insofar as it is so describable, that description will have to involve radically non-classical features, such as radical non-separability or the appeal to extremely high-dimensional spaces~\citep{albertmetaphysics,neyalbert}. 

\subsection*{Indefinite branch count}

How many worlds are there in the Everettian multiverse? The question is ill-defined~\citep[p.99]{wallacebook}. To describe quantum reality as a branching multiverse requires a choice of macroscopic variables, and that choice contains much arbitrariness: a description at one scale which says that there are 100 branches can be refined by a smaller-scale description which says that there are 10,000. Eventually the decoherence processes which make branches independent will break down, but there is no sharp point at which they do so; furthermore, there is no reason to expect that ratios of branch number are even approximately stable under fine-graining of this kind. Quantum mechanics tells us that there are many worlds, but it also tells us that there is not even an approximate fact of the matter as to how many worlds.

\subsection*{Trans-temporal identity}

Is the macroscopic world constantly splitting into more copies? Or should we think of worlds as four-dimensional entities, Big-Bang-to-Big-Crunch, whose temporal parts overlap at earlier times but which are themselves unbranching? The matter is controversial\footnote{See, \egc, \citep{saunderswallace,wilsonqmbook}).} and it is not entirely obvious that there is a fact of the matter; these might be alternative descriptions of the same underlying reality, a case of metaphysical indeterminacy~\citep[ch.7]{wallacebook}. But in any case, quantum mechanics tells us that at a sufficiently deep level, the apparently science-fictional, even pathological examples of temporally branching entities discussed by authors like~\citet{parfitbook} and \citet{lewisidentity} are in fact ubiquitous.

\subsection*{Self-locating uncertainty}

Philosophers have long argued\footnote{See, \egc, \citep{lewisindexical,perryindexical}} that in addition to beliefs about the objective truths of the universe, we can have \emph{self-locating} beliefs, beliefs about our location within a given possible world. But in pre-quantum physics this self-locating uncertainty is extremely rare. In a quantum universe it is ubiquitous: indeed, it seems likely that almost all of our beliefs, save only those about fundamental physics, are beliefs about our location in a branching reality where essentially all physically possible outcomes occur in some branch or other.

\subsection*{Probability as branch weight}

Science seems to require objective probabilities: perhaps ``Candidate X has a 30\% chance of winning the election'' just expresses a personal belief, but ``This neutron has a 50\% chance of decaying in the next 608 seconds'' seems to say something more robust. But the nature of those objective probabilities remains contested and obscure: proposals that they are relative frequencies, or axioms in the `best system' for describing the objective facts, or encodings of symmetry properties, or new metaphysical primitives, all have their respective adherents.\footnote{See \citep[ch.4]{wallacebook}, and references therein, for a review of these ideas in the quantum context.}

 If quantum mechanics is correct, then objective probabilities are (or at any rate supervene on) the squared moduli of quantum-mechanical amplitudes in the decoherent limit. This has the potential to transform our understanding of the metaphysics and epistemology of probability: for instance, it offers the possibility of deriving the `principal principle'~\citep{lewischance} that connects objective probability with personal credence from much weaker assumptions than are required in the classical context.

\section{Conclusion; or: what are we even doing here?}\label{section8}

The previous section was intentionally provocative. The vast majority of discussions of the philosophy of Everettian quantum mechanics do not presuppose the emergent multiverse and see what follows for our epistemology and metaphysics; rather, those discussions presuppose \emph{a priori} facts about epistemology and/or metaphysics, and use them to assess whether the Everett intepretation should be accepted or (more usually) rejected as philosophically unacceptable.\footnote{For instance: \citep{maudlinconference} on the micro-ontology of Everettian quantum mechanics; \citep{HemmoShenker2022} on the emergence of macro-worlds; \citep{adlamconfirmation,lewissu} on epistemology; \citep{albertinevbook,priceinevbook} on probability.}

And of course that is perfectly reasonable logically. If a priori philosophical view $X$ entails the falsity of Everettian quantum mechanics, and if I believe $X$, I cannot consistently also believe Everettian quantum mechanics. By analogy, if I hold a certain view on the metaphysics of temporal passage, and that view contradicts the special theory of relativity, I am committed to rejecting relativity. 

But (as the analogy suggests) this is not normally how we think about the philosophy of science. Generally we take scientific facts to constrain metaphysics and epistemology, and not vice versa. Why should quantum mechanics be any different? 

That isn't a rhetorical question. Concern about philosophical problems with the Everett interpretation might lead us to interrogate the story I have laid out here for the Everett interpretation and look for weak points in it. Possible options include:
\begin{enumerate}
\item I argued that given the basic structure of modern unitary quantum mechanics --- collapse-free dynamics, an emergentist story about the classical regime, based on the process of decoherence, a deflationary approach to `measurement' and `observers', a view that quantum mechanics is sufficient in principle to describe pretty much any physical process --- the Everett interpretation is unavoidable. That can be contested. Certain versions of the `relationalist' approach of Carlo Rovelli \emph{et al} (see, \egc, \citep{rovelli-relational}) can be understood as offering alternative readings of unitary  quantum mechanics, at the cost of abandoning a third-person, non-perspectival conception of physics. A sufficiently radical antireductionism could accept the emergent macroscopic description given by quantum mechanics as accurate while rejecting its microscopic foundations. Perhaps the long-neglected strategy of abandoning classical logic \citep{putnamquantumlogic,dicksonlogic,bacciagaluppiquantumlogic} could be resurrected. 

Any such strategies would themselves require a fairly radical rethinking of metaphysics and scientific interpretation, away from the broadly reductionist, objective-realist assumptions that underpin most of the metaphysics of physics (though, perhaps, closer to the more antireductionist approaches that philosophers like \citet{dupredisunity} and \citet{cartwrightdappled} advocate).

\item I asked you to take on trust my presentation of what modern quantum mechanics actually looks like. I argue for that presentation elsewhere~\citep{wallaceorthodoxy}, but if in fact this is not the correct rational reconstruction of contemporary physics, things might change.

\item Perhaps scientific reasons, theoretical and/or empirical (for instance, the Diosi/Penrose argument mentioned in section~\ref{section5}), might lead to the rejection of quantum mechanics wholesale. Its empirical success is so thoroughgoing that it has to remain accurate in large domains, but those might not include the domain of truly macroscopic superpositions. 
\end{enumerate}
But much\footnote{For instance: \citep{albertqmbook,maudlinqmbook,adlamsavingscience}} of the discussion of the Everett interpretation seems to have a different structure: it concedes that unmodified quantum mechanics is pretty much the Everett interpretation, thereby it rejects unmodified quantum mechanics on purely philosophical grounds, and so commits to the necessity of a scientific replacement for unitary quantum mechanics on \emph{a priori} philosophical grounds, even in the absence both of mathematical evidence that any such theory exists beyond the nonrelativistic regime, and empirical evidence of the limitations of the quantum framework.

To those philosophers who adopt this approach, all I can say is: I admire your confidence in the deliverances of analytic philosophy. To those philosophers who share my rather more jaundiced view of the track record for abandoning successful physics theories on the grounds that they conflict with our philosophical presuppositions: I recommend the exciting and wide open project of seeing how those philosophical presuppositions must change if we accept --- if only for the sake of argument --- that we live in the quantum multiverse.


\end{document}